\documentclass[12pt]{article}

\def\title{Not a prescription but an identity: Mandelstam-Leibbrandt.}

\long\def\abstract{
After some considerations and coincidences that appear when working
in the light-cone gauge, in both the Mandelstam-Leibbrandt prescription
and the covariantization method, we suspect that there must be some
connection between them. This work shows that we were right and it is 
practically trivial to demonstrate that relationship. 
And since the covariantization
method is not a prescription, this implies that the results of Mandelstam
and Leibbrandt are not prescriptions too, but they are a identities from
light-cone gauge coordinates.
\\
\\
{\bf PACS: 02.90.+p; 12.38.Bx}}

\newcommand{\intq}{\int{d^{2\omega}q}\cdot}
\input{axodraw.sty}
\input{epsf}

\begin{document}

\vskip 15mm
\centerline{\huge\bf Not a prescription but an identity:}
\centerline{\huge\bf Mandelstam-Leibbrandt.}
\vskip 8mm
\begin{center} \vskip 10mm
Ricardo Bent\'{\i}n{\begingroup\def\thefootnote{z}
                                \footnote{e-mail: rbentin@unifei.edu.br}
                                \addtocounter{footnote}{-1} \endgroup}
\\[3mm] Instituto de Ci\^encias, Universidade Federal de Itajub\'a - UNIFEI.\\
        Avenida BPS 1303, CEP 37500-903, Itajub\'a, MG, Brasil.
        
\vskip 3.0cm              {\bf ABSTRACT } \end{center}    \abstract

\thispagestyle{empty} \newpage
\pagestyle{plain} % \setcounter{page}{1} 

\newpage
\setcounter{page}{1}

\baselineskip 0.7cm
Leading with true degrees of freedom was one of the relishes of Dirac as
he showed us around in his famous paper ``On Hamiltonian Forms'' \cite{dirac}.
This should pointed out as the very beginning of light-cone gauge.
Since then, there were periods where people payed special attention on 
it, as an example, the case of quantization of the (super)string \cite{gs}.
But in quantum field theories some troubles appeared.
The use of light-cone gauge in (super)Yang-Mills theories ought to use
a prescription in order to avoid the pathological poles that arise 
on the propagator structure. The structure of this kind of poles 
brings the theory upon problems, just to mention three of them: 
Wick rotation is not allowed, naive power counting cannot be used and
non physical divergences are obtained from computing Feynman integrals.
The form of this poles, also known in the context of non covariant gauges
as spurious poles, is
$$
  \frac{1}{q.n}=\frac{1}{q^+},
$$ 
where $q$ is a vector of Minkowski spacetime of signature $(+,-,-,-)$ and
$n$ is a light-cone vector.\\
The use of the Principal Value (PV) prescription in Feynman integrals 
leads to not desire results, so another prescriptions were invented.
The first that appear was proposed by S. Mandelstam in 1983 when he was
working on $N=4$ super Yang-Mills theories. One year later, G. Leibbrandt
proposed another one, now in the context of Feynman integral calculations.
These prescriptions were in fact one \cite{bass}. But more than solve the 
problems cited above, and many others, the Mandelstam-Leibbrandt (ML)
prescription preserves causality, and this seems to be the mandatory
property that all well behaved prescriptions \cite{ps} must have on the 
light-cone gauge. An interesting work on the studies of symmetries of 
poles in the axial gauge is found in reference \cite{gaigg}.\\
One question may arise now, is a prescription really needed, mandatory,
or maybe there are other ways to obtain the same results but with 
prescriptionless methods?. Going in this way, A. Suzuki {\it et al}
\cite{us} found another exit.  
But there is also another technique, coined by A. Suzuki \cite{cov} as
covariantization, to treat light-cone integrals. This technique is
causal and reproduces the results obtained through the use of the
Mandelstam-Leibbrandt prescription \cite{ml}. 
This fact remains almost
forgotten, even though the covariantization manifestly guarantees the
absence of zero mode frequencies that spoil causality. Of course, here
the technique uses parametric integrations and integration over
components, and all the technology of complex analysis to treat the
singularities in a proper way. The thrust of the technique lies in
that it does not require a prescription for the light-cone pole; it
``converts'' this pole in a ``covariant'' pole whose treatment is
well-grounded and established since the early days of quantum field
theory. The burden of this technique and its most severe drawback is
that it requires an additional parametric integration to be performed,
a task which can be very demanding.

\section{Light-cone coordinates.}
Since our special purpose is to work with the light-cone gauge in
quantum field theory, then it is necessary to start, in our case very
briefly, with an introduction to light-cone coordinates. Using a 
signature $(+,-,-,-)$, then a general contravariant four-vector 
is given by the coordinates:
$$
  x^{\mu}=(x^+,x^-,x^i) \ ,   \ i=1,2,
$$
where $x^{\pm}$ are defined respect to usual Minkowski coordinates as:
\begin{eqnarray*}
    x^{\pm} &=& {\frac{1}{\sqrt{2}}}(x^0{\pm}x^3), \\
                &=&{\frac{1}{\sqrt{2}}}(x_0{\mp}x_3)=x_{\mp},\\
  x^i&{\equiv}&(\hat{x})^i =(x^1,x^2)
\end{eqnarray*}
Now, defining the dual base light-like four-vectors:
\begin{eqnarray}
  \nonumber
  n_{\mu} &=& {\frac{1}{\sqrt{2}}}(1,0,0,1),\\
  m_{\mu} &=& {\frac{1}{\sqrt{2}}}(1,0,0,-1),
\end{eqnarray}
we observe that with the help of this base, the $x^{\pm}$ coordinates
can be expressed as
\begin{eqnarray*}
        x^+ &=& x^{\mu}n_{\mu},\\
        x^- &=& x^{\mu}m_{\mu}.
\end{eqnarray*}
Using this coordinates, the scalar product becomes:
$$ 
x^{\mu}y_{\mu}=x^+y^-+x^-y^+-{\hat{x}}{\hat{y}},
$$
i.e.,
\begin{equation}
  \label{lcc}
  x^2=2x^+x^--{\hat{x}}^2,
\end{equation}
we will use this last expression later.
Also we have that the metric is now:
\begin{eqnarray*}
  g^{{\mu}{\nu}}&=&       \begin{array}{cc}
                                \begin{array}{cccc}
                                ^+ & ^- & ^1 & ^2
                                \end{array}
                                & \\
                                \left(
                                        \begin{array}{cccc}
                                                0 & 1 & 0 & 0 \\
                                                1 & 0 & 0 & 0 \\
                                                0 & 0 & 1 & 0 \\
                                                0 & 0 & 0 & 1\\
                                        \end{array}
                                \right)
                                &
                                \begin{array}{c}
                                ^+ \\ ^- \\ ^1 \\ ^2
                                \end{array}
                                \end{array}
\end{eqnarray*}
Now we are ready to go on.
 
\section{\large The ML prescription.}
In this section we will resume the ML prescription, showing how the Wick 
rotation is modified for the $(q.n)^{-1}$ poles when using this causal 
prescription.\\
First, the prescription reads
\begin{equation}
  \label{MaLe}
  \frac{1}{q^+}=\frac{1}{q.n}\rightarrow \frac{q^-}{q^+q^--\epsilon}.
\end{equation}
As said above, this prescription was first proposed by S. Mandelstam and
one year after, in 1983  by G. Leibbrandt. One of the important issues
this prescription takes on is the fact that Wick rotation can be done. 
\subsection{Wick rotation.}
\begin{figure}[h]
\vskip 0.5in
\centerline{
\begin{picture}(120, 170)(-40, -80)
\put(185, 5){\makebox(0,0)[br]{$Re(q^0)$}}
\put(-5,125){\makebox(0,0)[br]{$Im(q^0)$}}
\put(100,15){\makebox(0,0)[br]{${\cal C}$}}
\LongArrow(-170,0)(170, 0)
\LongArrow(170,0)(-170,0)
\LongArrow(0,-80)(0,140)
\LongArrow(0,140)(0,-80)
\ArrowLine(-110,10)(0,10)
\ArrowLine(0,10)(110,10)
\ArrowArc(0,10)(110,0,90)
\ArrowArc(0,10)(110,90,180)
\put(-95,20){\makebox(0,0)[br]{$\clubsuit$}}
\put(95,-20){\makebox(0,0)[br]{$\clubsuit$}}
\end{picture}
}
\caption{\label{covpoles}Poles $(q^2)^{-1}$ on the complex plane $q^0$.}
\end{figure}
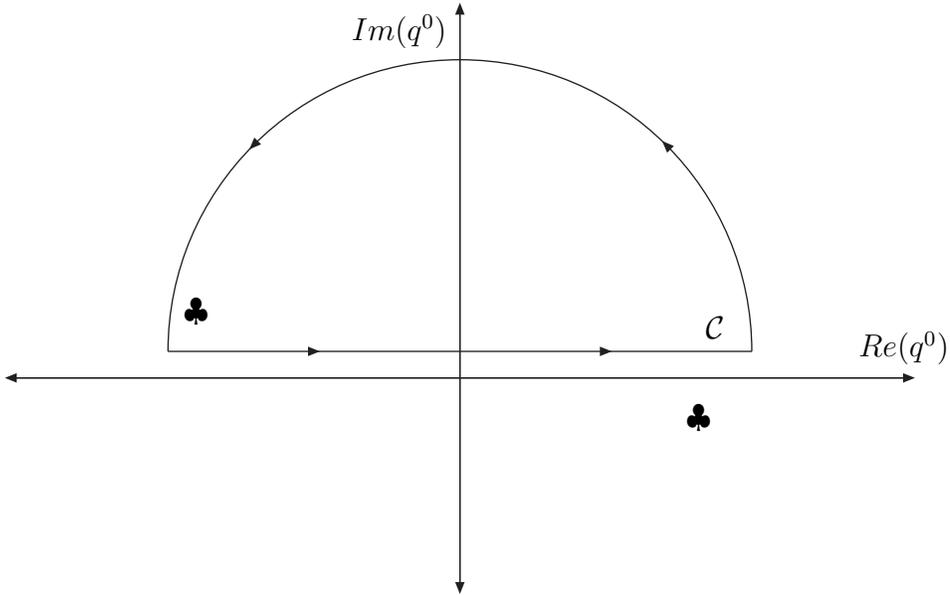

\begin{figure}[h]
\vskip 0.5in
\centerline{
\begin{picture}(120,150)(-40,-80)
\put(185,5){\makebox(0,0)[br]{$Re(q^0)$}}
\put(0,85){\makebox(0,0)[br]{$Im(q^0)$}}
\LongArrow(-170,0)(170,0)
\LongArrow(170,0)(-170,0)
\LongArrow(0,-80)(0,80)
\LongArrow(0,80)(0,-80)
\put(-95,20){\makebox(0,0)[br]{$\heartsuit$}}
\put(95,-20){\makebox(0,0)[br]{$\heartsuit$}}
\put(-85,+20){\makebox(0,0)[br]{$\spadesuit$}}
\put(-85,-20){\makebox(0,0)[br]{$\spadesuit$}}
\end{picture}
}
\caption{\label{lcgpoles} Light-cone gauge poles in the complex plane $q^0$. 
$\heartsuit$ arise from using ML prescription, $\spadesuit$ do not have a
prescription.}
\end{figure}
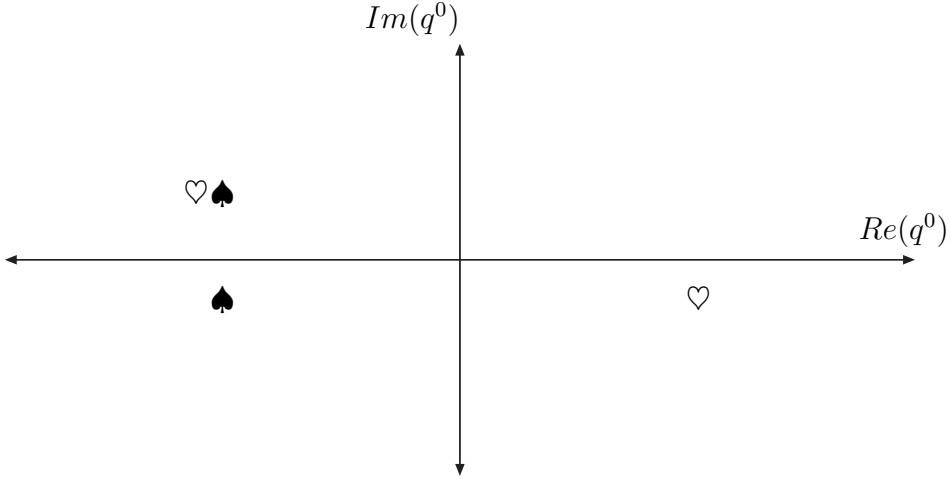

If one decides to use the path integral formulation of quantum field theory,
one of the problems we have to work out is the fact that the path integral
is no well defined in Minkowski spacetime. This is a delicate matter that 
belongs to a field called axiomatic field theory and it is known as 
Wick rotation.\\
First, we will use this rotation in working out the covariant poles of type 
$$
  \frac{1}{q^2},
$$
this is shown in figure \ref{covpoles}
%
%  Covariant pole
%
\vskip 0.5 cm
\vskip 0.5 cm
When working with the light-cone gauge, a special care must be taken since
poles of the type
$$
  \frac{1}{q.n}
$$
do not allow a Wick rotation if no causal prescription (e.g. ML prescription)
is used. This is shown in figure \ref{lcgpoles} where both cases are 
considered: with and without the use of the ML prescription. Observe the
location of poles in this two cases. This means that if we do not take care 
of use the ML prescription (or any other causal prescription) the result that
we will obtain are axiomatically wrong since we are working as if we were in
an Euclidean space when in fact we never leave the Minkowski spacetime.
This explains why all the pathologies are observed when the light-cone gauge 
is worked in the same manner of the other gauges. So we have a key point here,
the study of Wick rotation in the light-cone gauge is crucial and should be
considered as mandatory. 
%
%    $(q.n)^{-1}$ poles.
%

\subsection{Feynman integrals.}
To see how this prescription works, we will take the following two integrals
\begin{equation}
  \label{intA}
  A=\intq\frac{1}{(q-p)^2q^+}
\end{equation}
and
\begin{equation}
  \label{intB}
  B=\intq\frac{1}{q^2(q-p)^2q^+}
\end{equation}
as an examples. We are taking $D=2\omega$ as the dimension of spacetime,
i.e.; the limit to four dimensions means $\omega\rightarrow 2$.
%
%  A
%
Using the ML prescription given in eq. (\ref{MaLe}) and replacing in the 
first integral (A) of eq. (\ref{intA}), we obtain
$$
  A_{ML}=\intq\frac{q^-}{(q-p)^2(q^+q^--\epsilon)},
$$
the subscript $ML$ remarks that we are using the ML prescription.
At this point we have the chance of choosing a parametrization for the
denominator, we will use the exponential parametrization (also known as 
Schwinger parametrization, as you can see in the appendix), then
    \begin{eqnarray*}
      A_{ML}&=&-\intq q^-\int\limits^\infty_0d\alpha d\beta\cdot
          e^{i(\alpha(q-p)^2+\beta q^+q^-)},\\
         &=&-\intq q^-\int\limits^\infty_0d\alpha d\beta\cdot
          e^{i\alpha p^2}\cdot e^{i(2\alpha q^+q^--\alpha\hat{q}^2
            -2\alpha q^+p^--2\alpha q^-p^++2\alpha\hat{q}.\hat{p}
            +\beta q^+q^-)},
    \end{eqnarray*}
where we are working in light-cone coordinates. Interchanging the order of
the integrals we obtain
    \begin{eqnarray*}
      A_{ML}&=&-\int\limits^\infty_0d\alpha d\beta\cdot
          e^{i\alpha p^2}\cdot\intq q^-e^{i(2\alpha q^+q^--\alpha\hat{q}^2
            -2\alpha q^+p^--2\alpha q^-p^++2\alpha\hat{q}.\hat{p}
            +\beta q^+q^-)},\\
         &=&-\int\limits^\infty_0d\alpha d\beta\cdot
          e^{i\alpha p^2}\cdot\\
        &&\int d\hat{q}\cdot e^{-i(\alpha\hat{q}^2-2\alpha\hat{q}.\hat{p})}
          \int\limits^\infty_{-\infty}dq^-\cdot q^-e^{-2i\alpha q^-p^+}
          \int\limits^\infty_{-\infty}dq^+\cdot 
               e^{i(2\alpha q^--2\alpha p^-+\beta q^-)q^+}
    \end{eqnarray*}
now we are ready to work out the momentum integral. Starting with the $q^+$
component, it is nothing but the integral representation of Dirac distribution, 
this makes the $q^-$ integral trivial, and the $\hat{q}$ integral is of gaussian
type. After some work, the result obtained is
    \begin{eqnarray*}
      A_{ML}&=&-i^\omega\pi^\omega p^-\int\limits^\infty_0d\alpha d\beta\cdot
          \frac{\alpha^{2-\omega}}{(\alpha+\frac{\beta}{2})^2}
          e^{2i\frac{\alpha\beta}{2\alpha+\beta} p^+p^-},
    \end{eqnarray*} 
working the parameter integral: \footnote{observe that here we have made 
a rescaling $\beta$: $\beta\rightarrow 2\beta$.}
    \begin{equation}
      \label{Al}
      A_{ML}=i(-\pi)^\omega p^-(2 p^+p^-)^{\omega-1}
      \frac{\Gamma(2-\omega)\Gamma(\omega-1)}{\Gamma(\omega)}.
    \end{equation}
This is the final result for the integral $A_{ML}$.\\
Let's compute now the integral $B$, this will follow 
the same steps as above, then
%
%  B
%
$$
  B_{ML}=\intq\frac{q^-}{q^2(q-p)^2(q^+q^--\epsilon)},
$$
that means in this case,
    \begin{eqnarray*}
      B_{ML}&=&i\intq q^-\int\limits^\infty_0d\alpha d\beta d\gamma\cdot
          e^{i(\alpha(q-p)^2+\beta q^+q^-+\gamma q^2)},\\
         &=&i\intq q^-\cdot\\
         &&\int\limits^\infty_0d\alpha d\beta\cdot
          e^{i\alpha p^2}\cdot e^{ii[2(\alpha+\gamma)q^+q^-
            -(\alpha+\gamma)\hat{q}^2
            -2\alpha q^+p^--2\alpha q^-p^++2\alpha\hat{q}.\hat{p}
            +\beta q^+q^-]},
    \end{eqnarray*}
now we have 3 parameters
    \begin{eqnarray*}
      B_{ML}&=&i\int\limits^\infty_0d\alpha d\beta d\gamma\cdot
          e^{i\alpha p^2}\cdot\\
          &&\intq q^-e^{i[2(\alpha+\gamma)q^+q^-
            -(\alpha+\gamma)\hat{q}^2
            -2\alpha q^+p^--2\alpha q^-p^++2\alpha\hat{q}.\hat{p}
            +\beta q^+q^-]},\\
         &=&i\int\limits^\infty_0d\alpha d\beta d\gamma\cdot
          e^{i\alpha p^2}\cdot\\
        &&\int d\hat{q}\cdot e^{-i[(\alpha+\gamma)\hat{q}^2-2\alpha\hat{q}.\hat{p}]}
          \int\limits^\infty_{-\infty}dq^-\cdot q^-e^{-2i\alpha q^-p^+}\cdot\\
        &&\int\limits^\infty_{-\infty}dq^+\cdot 
          e^{i[2(\alpha+\gamma)q^--2\alpha p^-+\beta q^-]q^+}
    \end{eqnarray*}
    \begin{eqnarray*}
      B_{ML}&=&i^{\omega+1}\pi^\omega
         \int\limits^\infty_0d\alpha d\beta d\gamma\cdot
         \frac{\alpha}{(\alpha+\gamma/2)(\alpha+\beta+\gamma/2)^2}\\
        &&\cdot e^{i\alpha p^2+i(\alpha+\gamma/2)\hat{p}^2
           -2i{\frac{\alpha^2}{\alpha+\beta+\gamma/2}p^+p^-}},
    \end{eqnarray*}
finally
    \begin{equation}
      \label{Bl}
      B_{ML}=i(-\pi)^\omega p^-(p^2)^{\omega-2}
         \sum\limits^\infty_{n=0}
         \frac{\Gamma(3-\omega+n)\Gamma(\omega-1+n)}{\Gamma(\omega+n)\Gamma(n+2)}
         (1-\eta^{n+1})
    \end{equation}
where we defined
    $$\eta\equiv-\frac{\hat{p}^2}{p^2}.$$
These results of integrals (\ref{Al}) and (\ref{Bl}) are well known 
in the literature of non covariant gauges of axial type \cite{bass,leibb}.
For practically twenty years they are used as a base for other more
complicated results. 
\section{\large The covariantization method.}
Now we are going to show how the calculus of Feynman integrals that
present light-cone poles can be done without a prescription.
We will introduce the ``covariantization'' technique which was
proposed by A. Suzuki \cite{cov}. The main idea is quite simple. When
working with light-cone coordinates as was shown in eq. (\ref{lcc}), 
the square of a four-momentum is:
$$
  q^2=2q^+q^--\hat{q}^2,  
$$
As long as {\it $q^-\ne0$}, we can write $q^+$ as
$$
  q^+=\frac{q^2+\hat{q}^2}{2q^-}
$$
We note that this dispersion relation almost guarantees that real
gauge fields for which $q^2=0$ (real photons or real gluons for
example) are transverse; the residual gauge freedom, that is left to
be dealt with so that fields be manifestly transverse comes from the
presence of the $q^-$ in the denominator of the expression above.
This implies that in the light-cone gauge the characteristic pole 
becomes
\begin{equation}
  \label{cov}
  \frac{1}{q^+}=\frac{2q^-}{q^2+\hat{q}^2},
\end{equation}
The important thing here is that the condition $q^-\ne0$ warranties
the causal structure of this technique since it eliminates the
troublesome $q^-=0$ modes. Elimination of these modes restores the
physically acceptable results as can manifestly be seen in the causal
prescription \cite{ps} for the light-cone gauge.
\subsection{Wick rotation.}
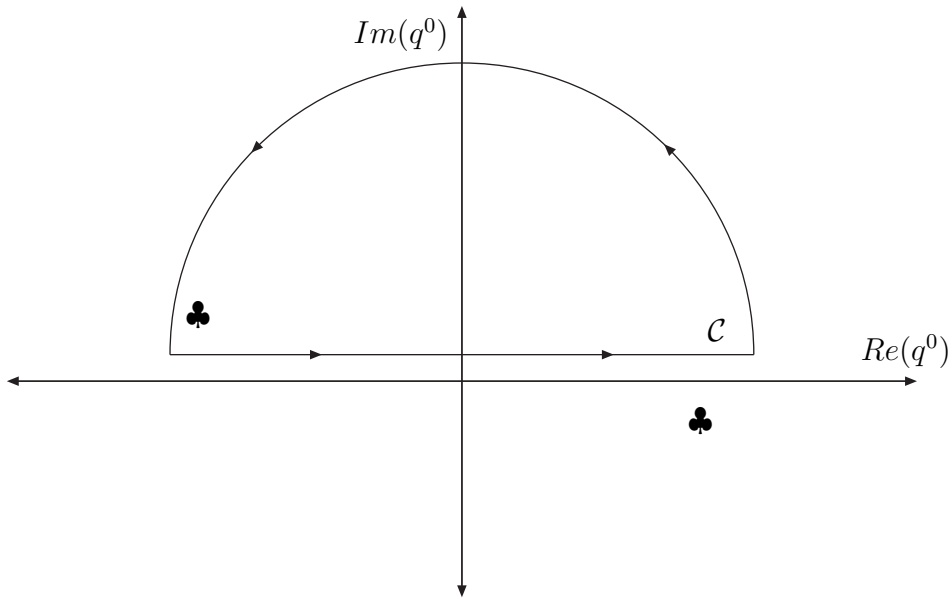
\begin{figure}[h]
\vskip 0.5in
\centerline{
\begin{picture}(120, 170)(-40, -80)
\put(185, 5){\makebox(0,0)[br]{$Re(q^0)$}}
\put(-5,125){\makebox(0,0)[br]{$Im(q^0)$}}
\put(100,15){\makebox(0,0)[br]{${\cal C}$}}
\LongArrow(-170,0)(170, 0)
\LongArrow(170,0)(-170,0)
\LongArrow(0,-80)(0,140)
\LongArrow(0,140)(0,-80)
\ArrowLine(-110,10)(0,10)
\ArrowLine(0,10)(110,10)
\ArrowArc(0,10)(110,0,90)
\ArrowArc(0,10)(110,90,180)
\put(-95,20){\makebox(0,0)[br]{$\clubsuit$}}
\put(95,-20){\makebox(0,0)[br]{$\clubsuit$}}
\end{picture}
}
\caption{\label{covP} Poles that arise when using the covariantization.}
\end{figure}

As pointed out in the above section, the importance of Wick rotation 
has physical implications in the light-cone gauge, and this will be an
important property that the covariantization method should have.\\
From light-cone coordinates, we obtained
$$
  \frac{1}{q^+}=\frac{2q^-}{q^2+\hat{q}^2}.
$$
This means that the denominator can be written as $(q^0)^2-(q^3)^2$,
making analytic continuation to the complex plane in the temporal 
component implies that the poles will be situated in
$$
  q^0=\pm|q^3|\mp i\epsilon
$$
as is shown in figure \ref{covP}.
%
%   Covariantization poles
%
\vskip 0.5 cm
\vskip 0.5 cm
This shows that the covariantization method also preserves the Wick 
rotation, which, when violated, was the cornerstone of non-physical 
results. If it is true, the results of integrals $A$ and $B$ will 
coincide when using covariantization.
\subsection{Feynman integrals.}
Taking again the integrals (\ref{intA}) e (\ref{intB}), but now with
the covariantization method, implies that
%
%   A
%
$$
  A_{cov}=\intq\frac{2q^-}{(q-p)^2(q^2+\hat{q}^2)}
$$
    \begin{eqnarray*}
      A_{cov}&=&-2\intq q^-\int\limits^\infty_0d\alpha d\beta\cdot
              e^{i(\alpha(q-p)^2+2\beta q^+q^-)}\\
           &=&-2\intq q^-\int\limits^\infty_0d\alpha d\beta\cdot
            e^{i\alpha p^2}\cdot e^{i(2\alpha q^+q^--\alpha\hat{q}^2
              -2\alpha q^+p^--2\alpha q^-p^++2\alpha\hat{q}.\hat{p}
              +2\beta q^+q^-)},
    \end{eqnarray*}
here we have used the relation: $q^2+\hat{q}^2 = 2q^+q^-$. Doing this
the appearance of $A_{cov}$ is closely similar to that of $A_{ML}$. 
This fact will help us when facing a possible relation between them.
    \begin{eqnarray*}
      A_{cov}&=&-2\int\limits^\infty_0d\alpha d\beta\cdot
          e^{i\alpha p^2}\cdot\intq q^-e^{i(2\alpha q^+q^--\alpha\hat{q}^2
            -2\alpha q^+p^--2\alpha q^-p^++2\alpha\hat{q}.\hat{p}
            +2\beta q^+q^-)},\\
         &=&-2\int\limits^\infty_0d\alpha d\beta\cdot
          e^{i\alpha p^2}\cdot\\
        &&\int d\hat{q}\cdot e^{-i(\alpha\hat{q}^2-2\alpha\hat{q}.\hat{p})}
          \int\limits^\infty_{-\infty}dq^-\cdot q^-e^{-2i\alpha q^-p^+}
          \int\limits^\infty_{-\infty}dq^+\cdot 
              e^{i(2\alpha q^--2\alpha p^-+2\beta q^-)q^+}
    \end{eqnarray*}    
following the same steps as before
    \begin{eqnarray*}
      A_{cov}&=&-2i^\omega\pi^\omega p^-\int\limits^\infty_0d\alpha d\beta\cdot
          \frac{\alpha^{2-\omega}}{(\alpha+\beta)^2}
          e^{2i\frac{\alpha\beta}{\alpha+\beta} p^+p^-},
    \end{eqnarray*} 
finally
    \begin{equation}
      \label{Ac}
      A_{cov}=i(-\pi)^\omega p^-(2 p^+p^-)^{\omega-1}
      \frac{\Gamma(2-\omega)\Gamma(\omega-1)}{\Gamma(\omega)},
    \end{equation}
which is the same result as obtained in eq. (\ref{Al}). This is in favor of
the covariantization method, but to remove any doubt, the same
computation will be done for integral $B$ and then compare it with 
the result of integral $B_{ML}$ of eq. (\ref{Bl}).
%
%   B
%
Then, we have
$$
  B_{cov}=\intq\frac{2q^-}{q^2(q-p)^2(q^2-\hat{q}^2)}
$$
again, following the same steps as before

    \begin{eqnarray*}
      B_{cov}&=&2i\intq q^-\int\limits^\infty_0d\alpha d\beta d\gamma\cdot
          e^{i(\alpha(q-p)^2+2\beta q^+q^-+\gamma q^2)},\\
         &=&2i\intq q^-\cdot\\
         &&\int\limits^\infty_0d\alpha d\beta d\gamma\cdot
          e^{i\alpha p^2}\cdot e^{i[2(\alpha+\gamma)q^+q^-
            -(\alpha+\gamma)\hat{q}^2
            -2\alpha q^+p^--2\alpha q^-p^++2\alpha\hat{q}.\hat{p}
            +2\beta q^+q^-]},
    \end{eqnarray*}
this implies
    \begin{eqnarray*}
      B_{cov}&=&2i\int\limits^\infty_0d\alpha d\beta d\gamma\cdot
          e^{i\alpha p^2}\cdot\\
          &&\intq q^-e^{i[2(\alpha+\gamma)q^+q^-
            -(\alpha+\gamma)\hat{q}^2
            -2\alpha q^+p^--2\alpha q^-p^++2\alpha\hat{q}.\hat{p}
            +2\beta q^+q^-]},\\
         &=&2i\int\limits^\infty_0d\alpha d\beta d\gamma\cdot
          e^{i\alpha p^2}\cdot\\
        &&\int d\hat{q}\cdot e^{-i[(\alpha+\gamma)\hat{q}^2-2\alpha\hat{q}.\hat{p}]}
          \int\limits^\infty_{-\infty}dq^-\cdot q^-e^{-2i\alpha q^-p^+}\cdot\\
        &&\int\limits^\infty_{-\infty}dq^+\cdot 
          e^{i[2(\alpha+\gamma)q^--2\alpha p^-+2\beta q^-]q^+}
    \end{eqnarray*}
doing the momentum integrals, 
    \begin{eqnarray*}
      B_{cov}&=&2i^{\omega+1}\pi^\omega
         \int\limits^\infty_0d\alpha d\beta d\gamma\cdot
         \frac{\alpha}{(\alpha+\gamma)(\alpha+\beta+\gamma)^2}\\
        &&\cdot e^{i\alpha p^2+i(\alpha+\gamma)\hat{p}^2
           -2i{\frac{\alpha^2}{\alpha+\beta+\gamma}p^+p^-}},
    \end{eqnarray*}
doing the parameter integrals,
    \begin{equation}
      \label{Bc}
      B_{cov}=i(-\pi)^\omega p^-(p^2)^{\omega-2}
         \sum\limits^\infty_{n=0}
         \frac{\Gamma(3-\omega+n)\Gamma(\omega-1+n)}{\Gamma(\omega+n)\Gamma(n+2)}
         (1-\eta^{n+1})
    \end{equation}
where again we defined
    $$\eta\equiv-\frac{\hat{p}^2}{p^2}.$$
this result agrees with that of $B_{ML}$.\\
This is not an ill-fated result. Instead, this makes us to 
think about the possibility of a certain kind of
relationship between the Mandelstam-Leibbrandt prescription and the
covariantization method. And that will be the motif of the following 
section.

\section{\large The ``proof''.}
This proof actually is not a real complicated proof, it is in fact
almost trivial as we can see now.\\
Let's take the form of a more general one loop Feynman integral in
the light-cone gauge,  
\begin{equation}
  \label{geral}
  I(p)=\int q.d^{2\omega}
       \frac{f(q,p)}{f_1(q,p)\cdots f_n(q,p)}\frac{1}{q^+}.
\end{equation}
Making use of the ML prescription, the integral (\ref{geral}) takes
the form
$$
  I_L(p)=\int q.d^{2\omega}
       \frac{f(q,p)}{f_1(q,p)\cdots f_n(q,p)}\frac{q^-}{q^+q^-+i\epsilon},
$$
after the Schwinger parametrization
\begin{equation}
  \label{gL}
  I_L(p)=
    (-i)^{n+1}\int q.d^{2\omega}\int\limits^\infty_0d\alpha_1\cdots d\alpha_nd\beta 
    \cdot q^-.e^{i[\sum\limits^n_1\alpha_if_i(q,p)+\beta q^+q^-]}.
\end{equation}
But if instead of using the prescription, we use the covariantization
method, we obtain
$$
  I_{cov}(p)=\int q.d^{2\omega}
       \frac{f(q,p)}{f_1(q,p)\cdots f_n(q,p)}\frac{2q^-}{q^2+\hat{q}^2},
$$
Schwinger parametrization implies 
\begin{equation}
  \label{gC}
  I_{cov}(p)=
    (-i)^{n+1}\int q.d^{2\omega}\int\limits^\infty_0d\alpha_1\cdots d\alpha_nd\beta 
    \cdot 2q^-.e^{i[\sum\limits^n_1\alpha_if_i(q,p)+\beta 2q^+q^-]},
\end{equation}
where the term that multiplies the $\beta$ parameter was changed using again 
the identity of light-cone coordinates
$q^2+\hat{q}^2\equiv 2q^+q^-$, 
in this way, the resemblance between the eqs. (\ref{gL}) e (\ref{gC})
is more evident. In fact, if the following change of variable 
is performed
\begin{eqnarray*}
  \beta &\rightarrow& 2\beta,\\
  d\beta &\rightarrow& 2d\beta,
\end{eqnarray*}
in eq. (\ref{gL}), this will coincide with that of eq. (\ref{gC}).
And also this is saying us that {\it the covariantization method and
the ML prescription are related} throughout a rescaling of coordinates
in the space of parameters.
\section{\large Conclusions.}
In this work we have shown that there exist a very close relationship 
between the wide used Mandelstam-Leibbrandt prescription and 
the covariantization method. Also, it is known that the covariantization
method is related \cite{covNDIM} to another technique, the negative
dimension integration (NDIM) for light-cone gauge \cite{us}, which is 
considered as a prescriptionless technique too. Since the covariantization
was just derived from a light-cone coordinates identity,
$$
  \frac{1}{q^+}=\frac{2q^-}{q^2+\hat{q}^2}\rightarrow\frac{q^-}{q^+q^-}, 
$$
this seems that something similar can be do for the ML prescription,
in this case,
$$
  \frac{1}{q^+}=\frac{q^-}{q^+q^--\epsilon}\rightarrow\frac{q^-}{q^+q^-}
$$
in both cases the same limit is reached, and in this limit the important
issue is the pole structure
$$
  \frac{1}{q^+q^-},
$$
which guarantees the Wick rotation and all well behaved results of the 
theory.\\
Summing up this work in few words, the ML prescription in fact should 
be called as the Mandelstam-Leibbrandt identity or the 
Mandelstam-Leibbrandt relation.\\

{\bf Acknowledgments:} We are grateful to professor A. T. Suzuki for
discussions on the covariantization method and professor R. Medina for
the hospitality at Universidade Federal de Itajub\'a.\\

\appendix
\section{Schwinger parametrization.}
The Schwinger parametrization consists on the following, starting from
the identity
$$
  \frac{1}{A}=-i\int\limits^{\infty}_0d\alpha\cdot e^{i\alpha A},
$$
it is possible generalize this to the case where we have $n$ denominators
\begin{equation}
  \label{Schwinger}
  \frac{1}{A_1A_2\cdots A_n}=(-i)^n\int\limits^{\infty}_0
    d\alpha_1d\alpha_2\cdots d\alpha_n e^{i\sum\limits^n_1\alpha_jA_j}
\end{equation}
in the case of working with the light-cone integrals, the form of the 
denominators are
\begin{eqnarray*}
        \frac{1}{(q-a_1)^2(q-a_2)^2{\cdots}(q-a_n)^2}
        \frac{1}{[[q^+]]}&=(-i)^n&
        \int\limits^{\infty}_0d{\alpha}_1d{\alpha}_2{\cdots}d{\alpha}_nd\beta\\
      &&e^{i[{\alpha}_iq^2-2{\alpha}_ia_i.q]}{\cdot}
        e^{i{\alpha}_ia^2_i}\cdot e^{i\beta[[q+]]}
\end{eqnarray*}
where $[[q^+]]$ stands for that we are using some causal prescription 
or the covariantization method.

\end{document}